\documentclass[acmsmall,screen]{acmart}

\usepackage{amsmath,amsfonts}
\usepackage{graphicx}
\usepackage{textcomp}
\usepackage{caption}
\usepackage{blindtext}
\usepackage{multicol}
\usepackage{xcolor}
\usepackage{tikz}
\usepackage{listings}
\usepackage{enumitem}
\usepackage{hyperref}
\usepackage{amsfonts}
\usepackage{wrapfig}
\usepackage{lipsum}
\usepackage{xspace} 
\usepackage{subcaption} 
\usepackage{adjustbox}
\usepackage{fancybox}

\usepackage[noend]{algpseudocode}
\usepackage[ampersand]{easylist}

\usepackage{colortbl}
\usepackage[normalem]{ulem}  
\useunder{\uline}{\ul}{}  
\usepackage{booktabs} 
\usepackage{multirow} 

\usepackage[T1]{fontenc}
\usepackage{tikz}
\usepackage{amsmath}
\usepackage{subcaption}
\usetikzlibrary{arrows.meta,positioning,calc}
\pgfdeclarelayer{bg}\pgfsetlayers{bg,main}   

\definecolor{causblue}{HTML}{1B6CA8}  
\definecolor{causgrey}{HTML}{C2C2C2}  
\definecolor{nodefill}{HTML}{F2F2F2}
\definecolor{nodeline}{HTML}{4D4D4D}
\definecolor{treatfill}{HTML}{D9E8F5}
\definecolor{outfill}{HTML}{E8DAF0}

\tikzset{
  latent/.style   = {circle, draw=black, line width=.7pt, fill=white,
                     minimum size=9mm, inner sep=0pt},
  treatnode/.style= {circle, draw=nodeline, line width=.7pt, fill=treatfill,
                     minimum size=9mm, inner sep=0pt},
  outnode/.style  = {circle, draw=nodeline, line width=.7pt, fill=outfill,
                     minimum size=9mm, inner sep=0pt},
  vnode/.style    = {circle, draw=nodeline, line width=.8pt, fill=nodefill,
                     minimum size=7.2mm, inner sep=0pt, font=\scriptsize},
  hnode/.style    = {circle, draw=nodeline, line width=1.3pt, fill=treatfill,
                     minimum size=7.2mm, inner sep=0pt, font=\scriptsize\bfseries},
  assumed/.style  = {draw=causblue, line width=1.5pt},
  other/.style    = {draw=causgrey, line width=.6pt},
  otherdir/.style = {draw=causgrey, line width=.9pt, -{Stealth[length=2.2mm]}},
  tally/.style    = {font=\scriptsize, text=causblue},
}

\newcommand{\ie}{\textit{i.e.,}\xspace}
\newcommand{\eg}{\textit{e.g.,}\xspace}

\newcommand{\etal}{et al.\xspace}

\newcommand{\tool}{{\sc{CauSec}}\xspace}
\newcommand{\tools}{{\sc{CauSec}}'s\xspace}

\newcommand{\cguard}{{CryptoGuard}\xspace}

\newcommand{\ccrypt}{{CogniCrypt}\xspace}

\newcommand{\sgrep}{{Semgrep}\xspace}

\newcommand{\cql}{{CodeQL}\xspace}

\newcommand{\detector}{{crypto-detector}\xspace}
\newcommand{\detectors}{{crypto-detectors}\xspace}

\usepackage[skins]{tcolorbox}

\newtcolorbox{takeaway}[1]{
    lower separated=false,
    colback=blue!2!white,
    colframe=teal!90!black,
    fonttitle=\sffamily\bfseries,
    left=1pt,
    right=1pt,
    bottom=1pt,
    top=4pt,
    colbacktitle=teal!90!black,
    coltitle=blue!2!white,
    enhanced,
    attach boxed title to top left={yshift=-0.1in,xshift=0.15in},
    boxed title style={boxrule=0pt,colframe=white,},
    title={\color{white}{#1}}
}

\newcounter{taskctr}

\newcommand{\myparagraph}[1]{\vspace{0.5em}\noindent{\bf #1:}}

\definecolor{dkgreen}{rgb}{0,0.6,0}
\definecolor{gray}{rgb}{0.5,0.5,0.5}
\definecolor{mauve}{rgb}{0.58,0,0.82}
\definecolor{shadecolor}{rgb}{0.95,0.95,0.95}
\setlist[itemize]{leftmargin=*, noitemsep, topsep=1pt}
\setlist[enumerate]{leftmargin=*, noitemsep, topsep=1pt}

\makeatletter

\DeclareGraphicsExtensions{.pdf,.png}
\graphicspath{{img/}}

\newcommand{\linebreakand}{%
  \end{@IEEEauthorhalign}
  \hfill\mbox{}\par
  \mbox{}\hfill\begin{@IEEEauthorhalign}
}

\definecolor{MidnightBlue}{HTML}{006895}

\newcommand\myparagraphnew[1]{\noindent{\bf {#1}:}}

\newcounter{fcounter}
\newcommand{\finding}[1]{\refstepcounter{fcounter}
\vspace{0.25em}\noindent\fbox{%
    \parbox{0.95\linewidth}{%
  \vspace{0.3em}{\bf
  {Finding~\arabic{fcounter}~(\fnumber{\arabic{fcounter}})}~--} {#1}
\vspace{0.3em}
  }
}
\vspace{0.25em}
}

\newcommand\fnumber[1]{{$\mathcal{F}_{#1}$}}
\newcommand\anumber[1]{{$\mathcal{A}_{#1}$}}

\newcommand{\boxme}[1]{{
\begin{tcolorbox}[enhanced,skin=enhancedmiddle,borderline={1mm}{0mm}{MidnightBlue}]
    \textbf{Finding } #1 \end{tcolorbox} 
}}

\newcommand{\scms}{SCMs\xspace}

\newcommand{\ate}{\textit{ATE}\xspace}

\newcommand{\sast}{{SAST}\xspace}
\newcommand{\sasts}{{SASTs}\xspace}

\AtBeginDocument{%
  }

\usepackage{xspace} 
\usepackage{amsmath}
\usepackage{floatflt}

\setcopyright{none}
\renewcommand{\footnotetextcopyrightpermission}[1]{}
\begin{document}

\title{CauSec: Unboxing the Causal Drivers of Static Analysis Security Testing Performance}

\author{Akram Khan}
\affiliation{%
  \institution{William \& Mary}
  \city{Williamsburg}
  \state{Virginia}
  \country{USA}
}

\author{Daniel Rodriguez-Cardenas}
\affiliation{%
  \institution{William \& Mary}
  \city{Williamsburg}
  \state{Virginia}
  \country{USA}
}

\author{Alejandro Velasco}
\affiliation{%
  \institution{William \& Mary}
  \city{Williamsburg}
  \state{Virginia}
  \country{USA}
}

\author{Denys Poshyvanyk}
\affiliation{%
  \institution{William \& Mary}
  \city{Williamsburg}
  \state{Virginia}
  \country{USA}
}

\author{Adwait Nadkarni}
\affiliation{%
  \institution{William \& Mary}
  \city{Williamsburg}
  \state{Virginia}
  \country{USA}
}

\renewcommand{\shortauthors}{Khan et al.}

\begin{abstract}
Static Application Security Testing (\sast) tools are widely used in industrial and academic software engineering. 
To balance vulnerability detection against practical objectives such as precision, runtime, and scalability, \sasts frequently make design decisions based on assumptions about the analyzed code or the analysis itself. 
These assumptions directly influence what vulnerabilities are ultimately detected, yet they are rarely evaluated using principled empirical methods. This raises a fundamental question: {\em are the assumptions underlying these design decisions actually valid?}

We address this question through the key observation that \sast assumptions are inherently causal in nature. 
We first conduct a systematic literature review of crypto-API misuse detectors, a security-critical \sast category, spanning the last two decades, identifying and characterizing 57 distinct security assumptions that vary widely in their objectives and amenability to causal evaluation.
Building on these observations, we present \tool, a framework for systematically evaluating SAST assumptions using causal inference. \tool{} formalizes security assumptions as causal hypotheses, operationalizes them as treatments, constructs structural causal models, estimates causal effects, and evaluates the robustness of the resulting conclusions.
We then conduct a comprehensive empirical causal evaluation of one representative assumption using four widely used crypto-API misuse detectors and a manually validated ground-truth dataset comprising 57,038 alerts. 
Our study shows that the validity and magnitude of security assumptions are fundamentally context dependent, varying across tools and library families, while also demonstrating both the strengths and limitations of causal inference as a methodology for evaluating SAST assumptions. 
Together, these findings provide new insight into both the design of SASTs and the principled evaluation of their underlying assumptions.

\end{abstract}

\authorsaddresses{}

\maketitle

\section{Introduction}
\label{sec:introduction}

Static Application Security Testing (\sast) tools play a critical role in preventing vulnerabilities in end-user software and are widely used in both industrial and academic software development workflows~\cite{Bessey2010,blackduckCoveritySAST,Wadhams2024}. 
They are routinely integrated into software quality assurance and continuous integration/continuous deployment (CI/CD) pipelines to  identify vulnerabilities before software release~\cite{xanitizer,githubGitHubShiftLeftSecuritysastscan}.
However, recent work has found that \sasts have {\em design and implementation flaws} that inhibit the detection of vulnerabilities~\cite{bkm+18,ack+22}, allowing them to persist in production software. 

Although flawed design choices could be implicit, \ie unintentional errors on the part of tool designers, some are well-intentioned attempts at addressing the trade-off between the detection rate and practical objectives such as precision and scalability. 
That is, \sast{}s often base their design choices on {\em assumptions} regarding code.
Consider \ccrypt~\cite{knr+17}, a popular \sast for detecting crypto-API vulnerabilities in software, or a \detector in short.
\ccrypt chooses not to report alerts from certain third-party libraries, as they could lead to increased false positives.
\cguard~\cite{rxa+19}, also a \detector, until recently, excluded alerts from core Android libraries for the same reason~\cite{ack+22,githubFixCryptoGuard}. 
Gaps in \detectors like \ccrypt can lead to highly vulnerable use of cryptography and hence severe consequences for consumer security and privacy~\cite{ack+22,mao+24}.
However, as seen in the examples above, assumptions shape the behavior of \detectors by motivating design decisions that determine what code is analyzed, what alerts are reported, and ultimately which vulnerabilities are detected.
Despite their practical impact, the validity of such assumptions is rarely established empirically. Instead, they are often accepted based on intuition, engineering experience, or limited observations, leaving it unclear whether the intended trade-offs actually materialize in practice. 
This raises a fundamental question: {\bf \em are the assumptions underlying gap-inducing design decisions in \sasts actually valid?}

This paper seeks to answer this question by contextualizing the approach of {\em causal inference}~\cite{Pearl2018Causality} to systematically test the assumptions made in \sasts. 
Our approach is motivated by a key observation: that the assumptions bounding the detection of \sasts are often {\em causal} in nature.
For example, the assumption that ``reporting alerts from third-party libraries leads to a decrease in precision'' implies a causal relationship between the condition (\ie reporting alerts from third-party libraries) and the outcome (\ie decreased precision). 
If formulated correctly, such assumptions can be expressed as causal hypotheses and tested via intervention.
With this intuition, this paper seeks to understand if, and how, causal inference might be used to shed light on the opaque assumptions in \sasts.
In particular, we investigate three research questions (RQs) that explore if and how causal inference is feasible and effective as a research methodology for understanding \sast assumptions:

\textbf{RQ1: {\em What do assumptions look like?}} What assumptions do tools make, what critical outcomes do they impact, and are all such assumptions amenable to causal analysis?

\textbf{RQ2: {\em How can we operationalize assumptions and validate them using causal inference?}} What representations, treatments, and causal models are needed to systematically evaluate \sast{} assumptions using causal inference.
%

\textbf{RQ3: {\em Can causal inference reliably validate \sast assumptions?}} How consistent and robust are the estimated causal effects across estimation methods, refutation tests, and data subgroups?

We scope the problem to understanding assumptions in \detectors (\eg, tools such as \ccrypt{} and \cguard{} described previously), given both the severe security consequences of their failures and the scale at which they are applied, which often motivates reliance on assumptions.
Our approach comprises two complementary components. 
To address {\bf RQ1}, we conduct a systematic literature review (SLR) of crypto-detectors spanning the last two decades and characterize the assumptions identified in the literature. 
To address {\bf RQ2} and {\bf RQ3}, we propose the \tool{} framework for systematically operationalizing and validating these assumptions using causal inference.
\tool proposes the abstraction of a {\em security assumption} that uniformly  captures the preconditions, conditions, and outcomes expressed in diverse assumptions found in \sasts. 
Building on this abstraction, \tool transforms informal security assumptions into explicit causal hypotheses, identifies interventions that operationalize them within the SAST workflow (\eg enabling or suppressing alerts from third-party libraries or changing exploration depth), and estimates their causal effects under an explicit causal model that accounts for confounding factors. 
By making the modeling choices explicit and applying refutation tests, \tool helps assess whether the resulting estimate remains stable under reasonable perturbations.
Thus, \tool enables assumptions that were previously accepted largely on intuition or engineering experience to be evaluated systematically using causal inference.

We evaluate \tool using a representative assumption that is prominently discussed in the literature on \detectors: that reporting alerts from third-party libraries leads to a decrease in precision.
Together, our systematic characterization of security assumptions and empirical evaluation of this representative assumption yield ten key findings that shed light on both the applicability of causal inference and the nature of assumptions underlying modern \detectors. 
This paper makes the following contributions:

\begin{itemize}
\item {\bf Characterizing Security Assumptions (the {\em what}) -- } We perform a systematic literature review (SLR) of \detectors\ {\em from the last 20 years}, which leads to the identification of 57 unique assumptions regarding diverse outcomes. 
We qualitatively analyze these $57$ assumptions to characterize recurring patterns and opportunities for causal analysis.

\item {\bf The \tool Framework (the  {\em how}) --} 
We design the \tool framework for evaluating \sast{} assumptions using causal inference. 
\tool formalizes the notion of a security assumption, enabling informal design rationales to be reformulated into operational forms suitable for causal evaluation.
It provides mechanisms for systematically testing these assumptions under explicit structural causal models.

\item {\bf Evaluation with \sast{}s (the {\em why}) --} We demonstrate the utility of \tool by evaluating a representative assumption \ie that reporting alerts from third-party libraries decreases precision, using four popular \sasts: \sgrep, \cql, \ccrypt and \cguard. 
Using a manually validated ground truth dataset of $57,038$ alerts, we estimate the adjusted causal effect of this assumption while controlling for measured confounders, and evaluate the effect for stability under various conditions and refutation tests.
\end{itemize}


Our empirical study of security assumptions and evaluation of \tool yield eleven findings (\fnumber{1}--\fnumber{11}), which we synthesize into four broader themes that illustrate the nature of assumptions, their amenability to causal inference-based evaluation, and their causal behavior.
These findings reveal previously unexplored relationships between design assumptions and detector behavior, demonstrating both the utility of causal inference as a methodology for studying software analysis tools, and the scientific insights such an analysis can provide. 


\section{Background}
\label{sec:background}

This section establishes the foundational concepts necessary for conducting a rigorous causal analysis of vulnerability analysis tools. We structure our section around the theoretical foundations using Pearl's framework~\cite{pearl2016causal,Pearl2009Causality,rodriguezcardenas2026rethinkingsoftwareempiricalstudies} that underpins causal reasoning and the practical methodologies for identifying and estimating causal effects from \sast assumptions as observational data.

Modern causal analysis is grounded in \textit{Structural Causal Models} (\scms)~\cite{palacio2024toward,Pearl2018Causality}, which encode causal relationships as structural equations over system variables. \scms are visualized as \textit{Directed Acyclic Graphs} (DAGs), where nodes represent variables and directed edges represent direct causal influence. A directed edge $T \to Y$ means that $T$ (treatment) appears in the mechanism that determines $Y$ (outcome). 

A {\em confounder} $Z$ is a variable that influences both $T$ and $Y$, introducing spurious associations that corrupt naive estimates of causal effects. Pearl formalizes this tension between \textit{seeing} (observational associations) and \textit{doing} (interventional effects), through \textit{do-calculus}~\cite{Pearl2018Causality,Pearl2009Causality,pearl2009overview}. Confounding is present when the observational and interventional distributions differ  \ie $p(y \mid t) \neq p(y \mid \text{do}(t))$.

To remove confounding bias, Pearl's \textit{backdoor criterion} identifies which variables $Z$ must be conditioned on to block all non-causal paths from $T$ to $Y$~\cite{pearl2016causal,neuberg2003causality,palacio2024toward}. Formally, $Z$ satisfies the backdoor criterion if it (1)~blocks all backdoor paths into $T$, and (2)~contains no descendant of $T$. When satisfied, the causal effect is identified by the backdoor adjustment:
\[
P(Y \mid \text{do}(T{=}t)) \;=\; \sum_z P(Y \mid T{=}t,\, Z{=}z)\; P(Z{=}z).
\]
This formula compares treated and control units \textit{within strata} of $Z$, then averages over the marginal distribution of $Z$, yielding an unbiased causal effect estimate.


\paragraph{Estimand, Estimator, and Estimate} Quantifying a causal effect requires answering three questions in sequence, (1) \textit{what} quantity should be estimated, (2) \textit{how} it should be computed from data, and (3) \textit{how much} the effect is. These correspond to the \textit{estimand}, the \textit{estimator}, and the \textit{estimate}.

The \textit{estimand} is the target causal quantity, expressed as a mathematical formula derived from the causal assumptions encoded in the DAG.
Common estimands include the Average Treatment Effect (\ate), the Average Treatment Effect on the Treated (\textit{ATT}), and the Conditional Average Treatment Effect \textit{(CATE)}, among others. 
In our study, we focus on the \ate, which measures the expected difference in outcomes between treated and control units across the entire population: $ATE = \mathbb{E}[Y(1) - Y(0)]$,
answering: \textit{on average, what is the effect for everyone?}

The \textit{estimator} is the methodological procedure used to compute the estimand from observed data, for example, propensity score matching (PSM), inverse probability weighting (IPW), or regression adjustment. 
The choice of estimator depends on the causal question, data characteristics, and modeling assumptions. Applying the estimator yields the \textit{estimate}, which is a numerical value of the causal effect accompanied by uncertainty quantifiers such as confidence intervals. 

\section{Related Work}

\tool builds on the theoretical foundations of
causal inference and its adaptations in software engineering, and to our knowledge, it is the first work to contextualize causal inference for the task of evaluating assumptions in SASTs. 
This section describes related work in two key areas:

\myparagraph{Causal Inference in SE} 
Causal reasoning is gaining traction in software engineering, with recent work applying Pearl's causal framework, including structural causal models, do-calculus~\cite{Pearl2018Causality,pearl2009overview}, and backdoor adjustment~\cite{pearl2016causal}, to problems such as interpreting neural code models~\cite{palacio2024toward,10336302} and studying trustworthiness of LLM-based code systems~\cite{smells_2026,ld_2026,sept_2026,palacio2024toward}.
However, prior work generally focuses on estimating causal effects for a specific software engineering problem once the treatment, outcome, and causal model are already defined.
In contrast, our work addresses a different challenge, as motivated by {\bf RQ1} -- {\bf RQ3}: what do design assumptions in security tools look like, and how to transform informal assumptions embedded in security tools into operational causal assumptions that can be systematically evaluated? 
This leads to the new abstraction of a {\em security assumption}, and its operationalization with the \tool framework for a reliable causal analysis. 

\myparagraph{Security Analysis Tools and Evaluations}
A large body of work has developed SASTs for detecting security vulnerabilities, including data leaks~\cite{arf+14,wror14,egc+10,gcec12,akg+15}, crypto-API misuse~\cite{fhm+12,rxa+19,SSG+14,Kruger2021,ebfk13,findsecbugsFindSecurity,blackduckCoveritySAST,sonarsourceCodeQuality, githubGitHubSecurity,githubGitHubShiftLeftSecuritysastscan}, and permission misuse~\cite{bkml14,skft09}. 
At the same time, there is growing recognition that these tools are often soundy in practice~\cite{lss+15}, making deliberate trade-offs in detection for gains in precision, runtime performance, or scalability.


While SASTs have been evaluated with benchmarks that often accompany the tools~\cite{ary19, lpp+22}, this realization has prompted dedicated research on evaluating these tools to uncover detection failures that violate the tools' claims. 
For instance, Bonett et al.~\cite{bkm+18} developed $\mu$se to evaluate data-leak detectors, while Ami et al.~\cite{ack+22} developed MASC to systematically evaluate crypto detectors.
These studies demonstrated that detection failures may arise not only from implementation defects but also from intentional design decisions and the assumptions that motivate them.

Our work is motivated by this line of research but differs in its objective. Rather than evaluating whether a SAST succeeds or fails on a benchmark, we focus on validating the underlying assumptions that drive its design choices. 
\tool is  the first effort toward systematically formalizing and operationalizing SAST assumptions through causal analysis, enabling us to estimate the effects of these assumptions in the context of their tools while explicitly accounting for confounding factors.

\section{Study of  Assumptions in Crypto Detectors (RQ1)}
\label{sec:assumption-analysis}
\tool is motivated by the assumptions in \sast{}s that influence their design decisions.
Extracting such assumptions from existing literature is essential to formulate them as security assumptions in \tool, so that they can be properly tested.
This section describes our methodology for extracting assumptions related to an important type of \sast, crypto-API misuse detectors, or {\em crypto detectors}, as well as the salient findings obtained from their analysis. 

\subsection{Methodology}
Security assumptions influence the design choices made by \sast tools and are, hence, very likely to be embedded in the literature describing \sast{}s, particularly research papers from the security and software engineering communities that propose \sast{}s.
Therefore, we leverage the systematic literature review (SLR) approach as our methodology for extracting assumptions, guided by standard best practices and recommendations from Kitchenham et al. \cite{kitchenham2009systematic}. 
Further, we qualitatively analyze the assumptions to uncover the latent insights and design rationale embedded in them.



\myparagraphnew{1. Identifying the information sources} 
We considered the proceedings of top-tier venues in security and software engineering (\eg~USENIX Security, CCS, S\&P, ACSAC, NDSS, ICSE, ASE, FSE), published in the last 20 years, and searched for relevant keywords (see the Appendix~\ref{app:assumptions}) in digital libraries, \ie the ACM Digital Library, IEEE Explore, and Google Scholar.  

\myparagraphnew{2. Inclusion and exclusion criteria} 
Our initial search led to a list of approximately $3000$ papers. 
To decide whether to consider a paper for further analysis, we applied the following {\bf \em inclusion criteria}: the paper is published in the last 20 years, and 2) crypto-API misuse detection as a goal.
We also applied the following {\bf \em exclusion criteria}: the paper is focused on dynamic analysis.
Following this methodology, we short-listed 20 papers for extracting assumptions (list in the artifact~\cite{anonymousgithub}). 

\myparagraphnew{3. Extracting Assumptions}
We extracted assumptions from the finalized papers using a single-coder approach, consistent with the prior studies~\cite{chen2024towards,ami2024false}.
An author read each paper completely and identified the assumptions, which included assumptions regarding the threat model, program modeling choices, analysis scope, coverage boundaries, and general precision, recall, and performance trade-offs.
The author recorded each assumption as a unique entry upon discovery and further analyzed other sections in each article to ensure correctness and consistency.
The broader list of assumptions was also gradually improved to ensure consistency and remove duplicates. 
To improve reliability, the extraction process included ongoing team feedback. 
The coder periodically shared the evolving assumption list with the team, and a second researcher reviewed the extracted assumptions for clarity and consistency.
For each assumption, we recorded metadata including the exact quote and its location in the paper. 
Through this process, we obtain $57$ unique assumptions. 

\myparagraphnew{4. Qualitative Analysis}
We used a single-coder qualitative coding approach to identify patterns across assumptions. 
Our coding followed an iterative development process. 
We started by creating a preliminary codebook based on our extracted assumptions, including the conditions (\eg presence of libraries) and outcomes (\eg precision, recall) in them, as well as the design choices and performance claims often discussed in static analysis tools (\eg program slicing, taint tracking, analysis depth). 
This initial codebook served as a foundation for our analysis, but evolved as we coded new assumptions in the data. 
After coding all the data, we iterated through the codes to uncover insights, which we discuss next.

\subsection{Assumption Analysis Results}
This section describes the results of our analysis of all $57$ assumptions, which we characterize along $5$ key themes. 
We support key observations with representative quotes from the studied papers and the associated assumption numbers (\anumber{i}) (see the artifact~\cite{anonymousgithub} 
 for the coded assumptions). 


\myparagraph{1. Strong claims regarding the causal effect of rules on detection accuracy}
While it is unsurprising that most if not all crypto detectors are rule-based systems, we observed that rule coverage is seen as the main factor that affects detection accuracy, which drives detector design as well.
%
For instance, several detectors focus on improving the ruleset as the primary approach to improving accuracy, \eg as evidenced by this statement in CryptoGo~\cite{li2022cryptogo}: {"\em To improve detection accuracy, we derive 12 cryptographic rules ..." }(\anumber{2}).
This statement illustrates an assumption about the ruleset (the condition) affecting accuracy (the outcome). 


Similarly, we observed tools making assumptions about the impact of compliance with their ruleset on the security of the analyzed software \eg {\em "We list the seven security rules that are used in our work, and {\bf \em any application} that violates any of those rules {\bf \em cannot be secure}"}(\anumber{39}). 
This instance highlights a particularly strong  assumption regarding the relationship between the ruleset and precision, \ie for the assumption to hold, certain rules would have to ensure 100\% precision, as the software ``cannot be secure'' if they are violated.
Moreover, detectors with different rulesets make similar assumptions regarding the impact of the rules on precision and accuracy, \eg the detector with 7 rules above~\cite{zhang2022example} considers them sufficient for absolute precision, while another detector with 12 rules~\cite{li2022cryptogo} claims the same.
This ambiguity regarding the real relationship between rules and detection accuracy  motivates our causal approach of evaluating such relationships. 



\finding{Detectors {\bf claim rules to be the main driver of detection accuracy}, with strong causal assumptions about their impact on recall and precision. 
} 

Since the notion of correctness is based on rules, detection performance can be sensitive to how the rules are configured.
We also find strong assumptions along these lines.
For instance, Semgrep*~\cite{bennett2024semgrep} (an enhancement to Semgrep~\cite{semgrep}) states that
{\em "By editing the configuration of Semgrep, we were able to increase its detection rate from 15.3\% (26/170) to 44.7\% (76/170)...
Configuring a single tool offers better detection rates and produces fewer warnings than using a suite of tools, as it does not produce duplicate warnings between tools."}~(\anumber{15}).
Here, Semgrep* makes a strong causal claim, that the causal effect of rule configuration on accuracy is {\em stronger than} that of ``using multiple tools''. 

\finding{Tools may make strong causal claims regarding deployment alternatives based on experimentally observed correlations, \eg claiming that a specific reconfigured tool has a greater impact on the detection accuracy than an ensemble of others. 
Such claims are highly likely to influence tool design and deployment.
}


\myparagraph{2. Focus on Precision-engineering}
A recurring observation is that \detectors assume a strong impact of pruning and refining on the precision of the analysis.
For example, CryptoGuard~\cite{rxa+19} makes two assumptions to this effect, \ie\ {\em "CryptoGuard addresses the false positive problem with a set of refinement algorithms derived from empirical observations of common programming idioms and language restrictions"}~(\anumber{7}) and {\em "Our detection streamlines the analysis by only analyzing what is necessary...Only after we slice through the import usage do we verify if there is a cryptographic misuse."}~(\anumber{12}). 
Similarly, some systems use slicing to filter out irrelevant statements to reduce noise, as seen in FireBugs~\cite{singleton2021firebugs}, \ie\ {\em "FireBugs performs intra-procedural program slicing to only include statements related to the API method of interest..."}~(\anumber{44}).

We find that precision is also assumed to be affected by how well the detector understands and maintains semantic context. 
Several assumptions suggest that warnings can be misleading when context is absent, such as when interprocedural effects are ignored. 
This leads to \emph{functional} false positives, where a pattern appears to indicate misuse but is not actually exploitable. 
As one detector aptly states: {\em "We show that without a proper contextual knowledge, the static program slicing approach, employed by...suffers from a significant ratio of functional false positives: a false positive that meets the  formal definition of misuse, yet does not introduce an exploitable vulnerability."}~(\anumber{26}).


\finding{Detectors treat pruning as an essential feature for reducing noise, and claim its strong causal effect on precision. Moreover, detectors also assume a strong association between retaining context and higher precision, considering ``functional'' false positives. }




\myparagraph{3. Trading accuracy for performance and scalability}
We find that tools repeatedly frame a broader goal of improving recall and detection accuracy, and yet, often reject or weaken it to avoid performance loss.
We observe several strategies and heuristics that represent unvalidated causal assumptions regarding performance gains, such as only focusing on portions of the control flow graph (\anumber{10}, from Cryptolation~\cite{frantz2024methods}), or only analyzing code on a per-file basis ( \anumber{34}). 

\myparagraph{4. Assumptions of Actionability}
The usefulness of the tool depends on whether the findings are actionable and whether the outcomes align with developer priorities.  
Tool designers recognize this, as we find several assumptions regarding a causal effect of certain design choices on actionability. 
Particularly, some tool designers believe that if tools show developers how to fix the vulnerabilities, their output would be more useful/actionable. 
For example, \emph{"FireBugs reports feedback to developers describing what is wrong with the implementation and how to fix it by showing correct usages of crypto APIs. ... integrate cryptographic components correctly and securely since they are often not cryptographic experts."} (\anumber{46}).
Similarly, tools also make assumptions about actionability that motivate them to implement additional techniques to diagnose and report the root cause of misuse (\anumber{40}). 


In contrast to such value additions, we observe that some assumptions regarding developer preferences also drive tools toward lowering the bar for detection.
Several tools assume that developers prefer an output with lower false positives, even at the cost of missing vulnerabilities, \eg as one states,  {"\em...we have learned that developers greatly prefer approaches that indicate actual vulnerabilities rapidly and with high precision over approaches that are sound but suffer from false positives and lack efficiency"} (\anumber{57}).
Such assumptions regarding usefulness, actionability, and developer preferences  are indeed causal in nature. 


\finding{Detectors assume that certain design add-ons (\eg incorporating fixes along with alerts) or optimizations (\eg prioritizing precision over recall) may have a causal impact on actionability and developer happiness. Testing these assumptions using causal analysis would be challenging as their outcomes must be measured in the field, and may be subjective.}

\myparagraph{5. Evaluation shapes the performance claims}
Finally, we observe that performance claims in tools can be misleading without clear definitions of the coverage and ground truth of the tool. 
To elaborate, we find assumptions supported by evidence that may highly dependent on the chosen scope of evaluation, \eg selecting a specific set of popular libraries: {\em "In practice, the filter utilizes the API data extracted from seven widely used open-source C/C++ cryptography libraries: libcrypto [42], libcrypt [29], cryptlib [4], LibTomCrypt [7], libgcrypt [6], wolfcrypt [9], and Nettle [8]. 
These libraries have covered nearly 100\% usage cases in our firmware dataset."} (\anumber{17}) 

Similarly, incomplete ground truth data can also negatively impact the validity of the evaluation. 
For example, as stated in Seader~\cite{zhang2022example}, \emph{"... the ground truth is incomplete, as it labels some instead of all invocations of Random().  We currently consider the extra calls of Random() found by Seader to be false positives, although the actual precision is higher." } (\anumber{37})

\finding{The measurements of precision, recall, and coverage in detectors may reflect how the test benchmarks were built, and {\bf do not sufficiently establish the causal effect} of the design decisions on the outcomes. }

\begin{figure*}[t]
  \centering
  \includegraphics[width=\textwidth]{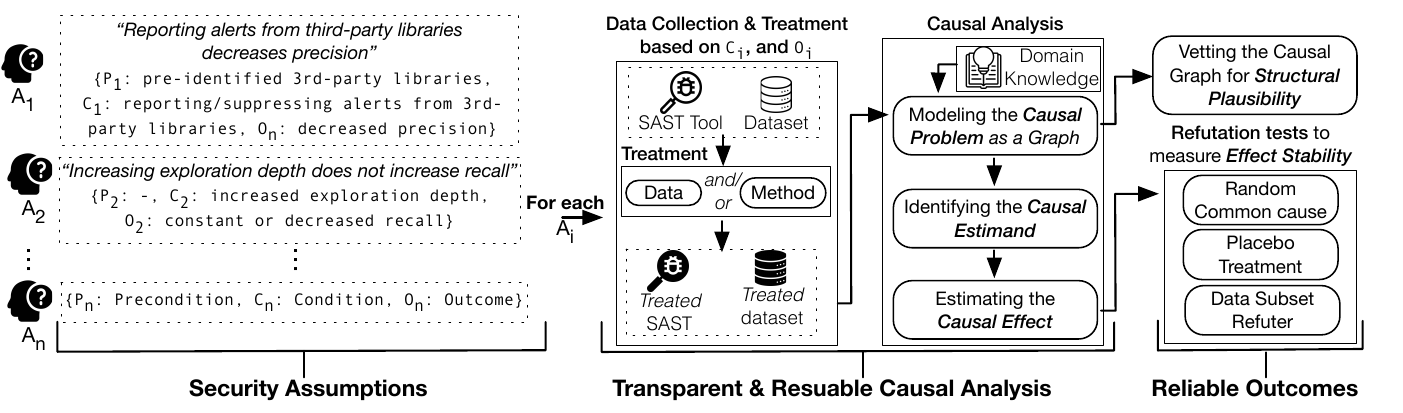}
  \caption{Overview of the \tool Framework}
  \label{fig:overview}
\end{figure*}

\section{The \tool Framework}
\label{sec:framework_design}

This paper introduces the \tool framework, which provides a systematic methodology for representing and evaluating the design assumptions made by \sasts{} using causal inference, thereby addressing RQ2, as illustrated in Figure~\ref{fig:overview}.

Security assumptions are often expressed informally and inconsistently across the documentation accompanying \sasts{}, including research papers, technical reports, and tool documentation. 
This lack of a common representation makes assumptions difficult to compare, reason about, and evaluate systematically.
\tool introduces the {\em security assumption}, which provides a precise structure to express assumptions using three key elements: a precondition , a condition ($C_i$), and an outcome ($O_i$).
This representation prepares assumptions for systematic causal evaluation with \tool, beginning with the identification of key factors and the preparation of data necessary for causal analysis.
In particular, condition ($C_i$) and outcome ($O_i$) help us identify the {\em treatment}, which may take the form of changes to the data, \sast methods, or, in some cases, both.
The outcome ($O_i$) also helps us identify the measurable behavior for causal analysis, such as precision, recall, or runtime. 

\tool evaluates a single security assumption at a time in the context of a single \sast. 
Given the formalized security assumption and the required data, \tool represents the causal inference problem as a structural causal model (SCM) that captures the relationships among treatments, outcomes, and potential confounders. 
\tool then identifies the causal quantity of interest (\ie the estimand), estimates its effect under the specified assumptions, and evaluates the robustness of the resulting estimate through refutation analyses (addressing both {\bf RQ2} and {\bf RQ3}). 
By making each modeling decision explicit and evaluating the stability of the estimated effects, \tool produces transparent, reproducible, and reliable causal evaluations of \sast assumptions.

\subsection{The Security Assumption}
\label{sec:security-causal-assumption}
Assumptions are often stated informally in artifacts and may not directly correspond to an intervention that can be evaluated through causal analysis. 
For example, consider the following assumption: ``Analyzing third-party libraries decreases precision''. 
Such assumptions about analysis behavior are often ultimately concerned with outcomes defined over the reported alerts, such as precision or recall.
We therefore reformulate informal assumptions into an operational form suitable for causal evaluation, preserving their intended meaning while making explicit the conditions and outcomes that define the causal hypothesis and guide the treatment. 
This reformulation determines {\em where} the treatment is applied during evaluation, such as at the analysis input, within the analysis process, or post-hoc on the analysis output.
For example, the informal assumption ``analyzing third-party libraries decreases precision'' can be reformulated as ``reporting alerts from third-party libraries decreases precision,'' (shown as $A_1$ in Figure~\ref{fig:overview}) since precision is measured over reported alerts rather than analyzed code. 
This reformulation exposes a post hoc treatment that modifies only the reported alerts, avoiding unnecessary changes to the application code or the \sast{} implementation while preserving the intended semantics of the original assumption. 
Many security assumptions can be operationalized similarly and evaluated with lower engineering and computational costs. 

To standardize assumptions in a consistent, understandable, and actionable form for causal analysis, we formulate the abstraction of a {\em security assumption}.
This abstraction captures the context in which an assumption applies, the condition it asserts, and the outcome it seeks to influence, thereby enabling the assumption to be operationalized as a causal treatment and evaluated through causal analysis.
%
A security assumption ($A_i$) is represented by a tuple consisting of the precondition ($P_i$),  the condition ($C_i$), and the outcome ($O_i$), as follows:
\[A_i = (P_i,\, C_i,\, O_i)\]
Here, the {\bf precondition} ($P_i$) specifies the scope or context in which the assumption is expected to hold. 
It captures any stated or implied prerequisites (\eg dataset constraints, platform constraints, or analysis configuration) that must be met for the assumption to be meaningful. 
Not all assumptions may be accompanied by preconditions. 
The {\bf condition} ($C_i$) is the property, heuristic, or constraint whose effect is being studied.
It forms the basis for constructing the causal treatment. 
The {\bf outcome}~($O_i$) describes the expected consequence once the condition is met, such as the impact on precision, recall, runtime, or scalability. 
Making the outcome explicit completes the formalization and provides an observable metric for estimating the causal effect associated with the assumption. 
Together, the condition $C_i$ and outcome $O_i$ determine how the treatment is constructed for causal evaluation.

As an example, consider an assumption we found in \textit{Cryptolation}~\cite{frantz2024methods}, a cryptographic API misuse detector for Python, as follows:
%
%
\textit{``Different from other SCA tools, we identify the imports for which we have misuse patterns and only scan for those. 
We call this an import-driven approach....{\bf \em Using an import-driven approach increases performance and does not sacrifice accuracy or precision}." 
} 
%
%
%
This statement captures not one but \underline{three distinct security assumptions}, because the same condition is associated with three different outcomes: improved runtime performance, a maintained detection rate, and maintained precision.
Across all three assumptions, the precondition ($P_i$) is that the analyzed program contains imports corresponding to cryptographic APIs for which misuse patterns are defined, and the condition ($C_i$) is that the analyzer adopts the import-driven strategy, restricting scanning and slicing to code regions associated with those imports.
The three assumptions differ only in their outcome ($O_i$). 

\subsection{Designing the Treatment}
\label{sec:treatment}

The security assumption also determines the data required for evaluation and how the treatment should be constructed from the condition and outcome.
Recall the assumption $A_1$ from Figure~\ref{fig:overview}, \ie that ``{\em Reporting  alerts from third-party libraries decreases precision.}''
To evaluate this assumption, we would need: {\sf (1)} a representative set of decompiled apps, {\sf (2)} identification of code paths in the apps that constitute third-party libraries, {\sf (3)} alerts produced by one or more \sasts (\eg \detectors in our evaluation) from analysis of the apps, and {\sf (4)} ground-truth true and false positive labels on all the alerts via manual validation.

As discussed in Section~\ref{sec:security-causal-assumption}, the reformulated security assumption determines where the treatment is applied within the \sast{} workflow (input, analysis process, or output).
For $A_1$, the treatment would be {\em reporting/considering alerts originating from third-party libraries} in addition to developer-origin alerts. 
As precision is defined over reported findings, the intervention is applied post hoc to the analysis output, yielding a data-level treatment. This avoids repeatedly executing the analyzer under different library configurations while preserving the semantics of the original assumption.
The intervention is thus applied post-hoc to the reported alerts, constituting a change to the alert dataset, \ie a {\em data-level} treatment.
This approach avoids repeatedly executing the \sast under different library configurations while preserving the semantics of the original assumption.


In contrast, if the outcome $O_1$ were {\em runtime} rather than precision, \eg if the assumption is that ``analyzing third party libraries significantly increases runtime'', a post hoc treatment would no longer preserve the semantics of the assumption.
Instead, the intervention must occur before or during analysis, by modifying the analyzed apps (data-level) or the analysis method (method-level treatment) to exclude/include libraries, so that the measured runtime reflects the treatment.

\subsection{Modeling the Causal Analysis Problem}
\label{sec:modeling-causal-problem}

After preparing the data and treatment, \tool represents the security assumption as a structural causal model (SCM), (Section~\ref{sec:background}), which specifies the direct dependencies among variables while allowing for unmeasured background factors. 
The model makes explicit the treatment, outcome, and potential confounding variables that influence both. 
We represent these relationships using a directed acyclic graph (DAG), which serves as the basis for causal identification, effect estimation, and robustness analysis.


\myparagraph{Unit of analysis and outcome}
For our running example, the unit of analysis is an alert, \ie a a single report produced by an \sast tool on a specific code location. 
The outcome variable is precision, which we treat as correct if manual validation confirms it as a true positive.
\begin{itemize}
    \item \textbf{Outcome ($Y$):} \textit{verdict} $\in \{0,1\}$, where $1$ denotes a true positive (TP) and $0$ a false positive (FP).
\end{itemize}
The estimated effect should therefore be interpreted as a change in the probability that a reported alert is correct, rather than as a deployment-level effect that also considers alert volume, deduplication, or developer triage.
\begin{floatingfigure}[l]{1.5in}
\centering
\resizebox{0.3\textwidth}{!}{%
\begin{tikzpicture}[>={Stealth[length=2.4mm]}, scale=0.95, transform shape]
  \useasboundingbox (-1.05,-2.15) rectangle (5.15,3.60);
  \node[latent]    (Z) at (2.15, 2.20) {$Z$};
  \node[treatnode] (T) at (0.45, 0.55) {$T$};
  \node[outnode]   (Y) at (3.85, 0.55) {$Y$};

  \node[font=\small]                at (2.15, 3.35) {Confounders};
  \node[font=\small\itshape]   at (1.00, 2.85) {\textless app\_popularity\textgreater};
  \node[font=\small\itshape]   at (3.45, 2.85) {\textless apk\_size\textgreater};

  \draw[dashed, ->, line width=.7pt] (Z) -| (T);
  \draw[->, line width=.7pt]         (Z) -| (Y);

  \draw[->, line width=.8pt] (T) --
        node[above=1pt, align=center]
          {$P(Y\,|\,do(T))$\\[-2pt]{\scriptsize\itshape ATE}} (Y);

  \node[below=1.5mm of T, align=center, font=\small] {Treatments\\{}[Binary]};
  \node[below=1.5mm of Y, align=center, font=\small] {Verdict\\Outcomes};

  \node[font=\small, align=center] at (2.05,-1.80)
        {$T_0$: Dev-only,\quad $T_1$: Incl. 3rd-party-lib\\[1pt]
         $Y_0$: TP,\quad\, $Y_1$: FP};
\end{tikzpicture}%
}
\caption{\small The SCM for evaluating $A_1$.}
\label{fig:scm}
\end{floatingfigure}

\myparagraph{The Structural causal model}
The SCM lets us reason about interventions explicitly by setting a treatment while holding the rest of the system fixed.
One such causal graph for our running example is shown in Figure~\ref{fig:scm}.
%
In the graph, T=the treatment is the reporting policy (\textit{3rd-party-lib}), which indicates whether alerts from third-party libraries are reported along with alerts from developer-only code. 
The outcome is the manually validated alert verdict, where 1 denotes a true positive and 0 a false positive.
We additionally model app popularity and APK size as contextual variables because they may influence both the treatment and the outcome.
App popularity, measured by download count, can influence dependency usage patterns~\cite{wang2017understanding}, thereby affecting both the prevalence of third-party library alerts and alert correctness through ecosystem factors such as code organization and dependency practices. 
Similarly, larger applications tend to contain more third-party code and more complex dependencies, making app size another plausible confounder.
Finally, the reporting policy influences alert correctness because alerts originating from third-party libraries may exhibit different true-positive and false-positive characteristics than alerts from developer-written code.
Directed edges in Figure~\ref{fig:scm} describe these confounding relationships.

This graph is deliberately compact, as we include only variables that are (i) measurable in our evaluation pipeline, and (ii) plausibly responsible for confounding or heterogeneity in alert correctness based on domain knowledge. 
Instead, it makes explicit the variables for which adjustment is possible, and the assumptions under which the estimated causal effect should be interpreted.

\subsection{Identifying the Causal Estimand}
\label{sec:identify-estimand}

After specifying the causal graph, \tool identifies the causal estimand, \ie the causal quantity representing the effect of the treatment on the outcome.
For a fixed causal graph and treatment-outcome pair, the same estimand defines the target causal quantity across datasets~\cite{Pearl2018Causality}.
In our setting, the outcome is the manually validated alert verdict, so the estimated effect is interpreted as the change in the probability that a reported alert is a true positive (i.e., precision).
We use the Average Treatment Effect (ATE), as our goal is to quantify the average effect implied by a security assumption across the population under study, \ie all applications in the dataset.

Defining an estimand is not sufficient; it must also be identifiable from observational data. 
Identification links the interventional quantity $P(Y=1 | do(T=t))$ to an expression that can be estimated from observed data. 
The causal graph plays a central role here by making potential confounding relationships between the treatment $T$ and the outcome $Y$ explicit.
%
%
%
%
\tool uses the backdoor criterion~\cite{pearl2016causal} as the identification strategy because its treatments are observational rather than randomly assigned. 
For example, code provenance, popularity tier, or library family may share common causes with alert correctness. 
The backdoor criterion identifies an adjustment set that blocks these non-causal paths, allowing the interventional effect to be estimated from observed data under the assumed causal model.

\subsection{Estimating the Causal Effect}
\label{sec:estimate-causal-effect}

After identifying the causal estimand, \tool estimates the causal effect from observational data using statistical and machine learning methods. 
For the running example, the treatment is binary, and therefore \tool uses propensity score matching (PSM) to estimate ATE.  
In our evaluation, we additionally apply logistic regression adjustment and Doubly Robust Estimation (AIPW) to assess whether the direction and magnitude of the estimated effect remain consistent across matching-based and model-based estimators.

Since the outcome $Y$ is binary (\textit{verdict} $\in \{0,1\}$), the ATE is directly interpretable as the change in the probability that a reported alert is a true positive:
\[
\text{ATE}=\mathbb{E}[Y\mid do(T=1)]-\mathbb{E}[Y\mid do(T=0)].
\]
\vspace{-1em}

\subsection{Robustness of \tools Outcomes}
\label{sec:validating-causal-process}
The final stage of \tool assesses robustness at two levels. 
First, we check whether the causal graph aligns with basic empirical patterns in the observed data. 
Second, we test whether the estimated effect remains stable under perturbations that should either preserve the effect or nullify it.
These checks help \tool identify 
estimates that are stable and therefore credible under the causal model.

\myparagraph{Vetting the causal graph using empirical checks}
We first assess the causal graph for {\em structural plausibility} by comparing key relationships implied by the graph against observable data.
%
For example, for edges in the graph that were suggested by domain knowledge, \eg such as app popularity or app size influencing the treatment in Figure~\ref{fig:scm}, we calculate simple association measures (\eg Spearman correlation) using the application and labeled alert datasets prepared for causal estimation. 
These checks do not establish causality, but help identify graph assumptions that are inconsistent with the observed data.
%

\myparagraph{Evaluating robustness of the estimated effect}
Refutation tests perturb the data or variables to assess {\em effect stability}. 
Rather than proving causality, these tests assess whether the estimated effect remains stable under perturbations that should not meaningfully alter a valid causal relationship.
\tool uses the following refuters: 

\begin{enumerate}[label=\textbf{R$_{\arabic*}$}, ref=\textbf{R$_{\arabic*}$}, wide, labelindent=5pt]\setlength{\itemsep}{0.2em}
    \item \textbf{Random common cause.}\label{ref:random}
    An independent random variable $R$ is added to the model as an additional common cause. 
    Because $R$ is unrelated to the treatment and outcome, a stable estimate should remain essentially unchanged.
    \item \textbf{Placebo treatment.}\label{ref:placebo}
    The original treatment is replaced with an independent placebo treatment of the same type, while the rest of the data remains fixed. 
    Since the placebo treatment should have no real causal connection to the outcome, the estimated effect should move close to zero.

    \item \textbf{Data subset.}\label{ref:subset}
    The effect is re-estimated on randomly selected data subsets. 
    A stable estimate should remain directionally consistent and close to the original estimate across these subsets.

    \item \textbf{Dummy outcome.}\label{ref:dummy}
    The original outcome is replaced with a synthetic one generated to have no causal dependence on the treatment. 
    A stable estimator should produce an effect close to zero for this dummy outcome. 
    \item \textbf{Unobserved Common Cause.}\label{ref:unobserved}
    A simulated unobserved variable is introduced as a common cause of the treatment and outcome. 
    The resulting change in the estimate indicates its sensitivity to omitted confounding.
\end{enumerate}

\section{Empirical Causal Study with \tool}
\label{sec:causec-evaluation}

We now conduct a thorough empirical causal evaluation of one representative assumption identified in Section~\ref{sec:assumption-analysis}.
To ensure the evaluation is empirically meaningful, we select \anumber{1} \ie that {\em ``reporting alerts from third-party library code decreases precision''}, as it {\sf (1)}
is common across multiple \sast{}s, {\sf (2)} can be evaluated using
observable outcomes, and {\sf (3)} can be tested without substantial
re-engineering of existing tools.
We operationalize this assumption as a reporting-policy treatment and examine whether the empirical evidence supports it across the four representative \detectors detectors, addressing {\bf RQ2}.
We then conduct a comprehensive robustness analysis of the estimated causal effects across multiple estimators, refutation tests, uncertainty specifications, and library categories, thereby addressing {\bf RQ3}. 

While the resulting causal estimates are specific to this setting (\ie a representative assumption, and 4 popular \detectors), the \tool framework is designed to apply to the broader class of security assumptions identified in our systematic study (Section~\ref{sec:assumption-analysis}). 
Thus, beyond evaluating this particular assumption, the study provides insight into {\em how} causal inference can be systematically applied to understand and validate \sast{} assumptions more broadly.
\subsection{Experimental Setup}
\label{sec:causec-experimental-setup}

We implemented \tool{} using the DoWhy library~\cite{sharma2020dowhyendtoendlibrarycausal}, which supports end-to-end causal analyses over structured datasets. 

\myparagraph{1. App sampling} For a representative analysis, we obtained a stratified sample of  APKs to analyze for crypto-API misuse, balanced across 11 popularity strata as given by their install count (\ie 100-500, 500-1k, 1k-5k, ...1M-5M, $>$5M installs).
To elaborate, we downloaded APKs from AndroZoo~\cite{allix2016androzoo} in June 2025, and decompiled them using jadx (version 1.5.2).
For each stratum, we sampled until we could obtain at least 50 APKs that would decompile, and were not obfuscated, leading to a final dataset of 558 APKs across 11 popularity strata. 

\myparagraph{2. Ground Truth Alert Dataset}
We analyzed the decompiled APKs using four widely-used \detectors, namely Semgrep\cite{semgrep}, CodeQL\cite{codeql}, CogniCrypt~\cite{knr+17} and CryptoGuard~\cite{rxa+19}.
These tools were chosen based on popularity, relevance to industry and academia, and causal statements regarding third-party libraries.
After removing alerts from partially decompiled or obfuscated code, we obtained  $57,038$ alerts ($14,590$ from \sgrep, $21,501$ from \cql, $5,675$ from \ccrypt and $15,272$ from \cguard).
One author manually validated these alerts to label them as true or false positives, using the alert description, code location, and the surrounding context, following a standard validation approach for similar large-scale studies~\cite{chen2024towards,ami2024false}.
This process took approximately 4.5 person-months and resulted in a {\bf ground-truth dataset of 57,038 labeled alerts}.


\subsection{Identification and Structural Robustness of the Causal Graph}
\label{sec:discovery-results}
\begin{floatingfigure}[l]{1.7in}
\centering
\resizebox{1.7in}{!}{%
\begin{tikzpicture}[scale=0.9, transform shape]
  \useasboundingbox (-0.70,-1.0) rectangle (4.70,3.60);
  \node[vnode] (ver)  at (-0.05, 1.45) {ver};
  \node[vnode] (util) at ( 0.95, 2.30) {util};
  \node[vnode] (n)    at ( 2.00, 1.70) {n};
  \node[vnode] (soc)  at ( 3.05, 2.30) {soc};
  \node[vnode] (andr) at ( 4.05, 1.45) {andr};
  \node[vnode] (size) at ( 1.05, 0.65) {size};
  \node[vnode] (pop)  at ( 2.95, 0.65) {pop};
  \node[hnode] (trd)  at ( 1.05,-0.65) {3rd};
  \node[hnode] (tp)   at ( 2.95,-0.65) {TP};

  \begin{pgfonlayer}{bg}
  \foreach \a/\b in {size/trd, size/tp, pop/trd, pop/tp}
      \draw[assumed, -{Stealth[length=2mm]}] (\a) -- (\b);
  \draw[assumed, -{Stealth[length=2mm]}] (trd) to[bend right=8] (tp);
  \end{pgfonlayer}
\end{tikzpicture}%
}
\caption{\small \tool causal graph. \footnotesize{ver= version code, TP=True Positives, Pop=App popularity, andr= Android, soc= social media library, util= utilities library.}}
\label{fig:causec-assumed-graph}
\end{floatingfigure}
We identified the causal graph iteratively, beginning with the simplest plausible specification and progressively introducing candidate adjustment variables based on domain knowledge and empirical sensitivity analyses.
%
We considered three candidate variables: app popularity, APK size, and rule identifier (\texttt{rule\_id}). 
App popularity and APK size are plausible confounders as they may affect both the treatment assignment and the outcome. 
In contrast, {\em rule\_id} was modeled as an {\em effect modifier} rather than a confounder, since it may influence alert precision but does not determine whether an alert originates from developer-written or third-party library code.
We evaluated candidate graphs using all combinations of the candidate confounders, with and without \texttt{rule\_id} as an effect modifier.
In all cases, adding  \texttt{rule\_id} did not change the estimated effect, indicating that the final estimate is not driven by rule-specific heterogeneity. 
We therefore selected the graph in Figure~\ref{fig:causec-assumed-graph}, which blocks the principal plausible backdoor paths while excluding \texttt{rule\_id} from the adjustment set.
This graph was consistently used throughout the causal estimation process.


In addition to the evaluation of candidate graphs above, we applied 4 causal discovery algorithms from Hulse \etal~\cite{hulse2025shakystructureswobblyworld} (GES, PC, FCI, and DirectLinGAM) that recover and suggest causal graphs from the dataset.
Across four causal discovery algorithms and four perturbation settings, we recovered 3,632 candidate graphs. 
%
%
The best of the recovered graphs, \ie the one with the most edges, discovered 4/5 edges from our causal graph in Figure~\ref{fig:causec-assumed-graph}.
However, in general, the recovered graphs exhibited substantial variability. 
In particular, \texttt{apk\_size} was adjacent to neither the treatment nor outcome in approximately 75\% of recovered graphs, while DirectLiNGAM rarely identified app popularity as a confounder.
Conversely, one variable outside the assumed adjustment set, \texttt{alerts\_in\_app}, appeared as a potential confounder in 42\% of the recovered graphs.
These observations suggest that causal discovery alone is insufficient for constructing the final causal graph. 
Instead, domain knowledge remains essential for selecting a plausible adjustment set, while causal discovery provides complementary evidence that can guide future refinements of the model.

\finding{Causal discovery complements, but does not replace, domain knowledge when constructing causal graphs for evaluating \sast{} assumptions.}

%
%
%

\subsection{Results: Estimating Causal Effects using \tool}
\label{sec:causec-results}
As a baseline prior to causal adjustments, we provide the raw precision of alerts overall and by alert source in the Table~\ref{tab:raw_precision_results}, and describe specific values here.   
We observe that including library-code alerts lowers raw precision for \cql, \ccrypt, and \cguard: from 0.95 to 0.89 for \cql, from 0.70 to 0.48 for \ccrypt, and from 0.51 to 0.48 for \cguard.
In contrast, \sgrep precision increases from 0.55 on alerts from developer-written code to 0.67 when library-code alerts are included.
As the datasets are highly imbalanced, with fewer alerts from developer-written code as opposed to those in third-party libraries, we present two simple balancing methods for all four datasets: (A) downsampling third-party alerts to match the number of developer-written alerts and (B) bootstrapping developer-written alerts to equal the number of third-party alerts. 
Both methods show the same difference seen in the unbalanced case. 
We use these summaries as observational baselines for subsequent causal analyses.

\begin{table}[t]
\centering
\caption{PSM causal estimates, APK-cluster bootstrap confidence intervals, and refutation results.}
\label{tab:overall_refuter_results}
\scriptsize
\scalebox{1}{
\begin{tabular*}{\linewidth}{@{\extracolsep{\fill}}
    >{\raggedright\arraybackslash}p{0.11\linewidth}
    c
    >{\centering\arraybackslash}p{0.16\linewidth}
    l r r r@{}}
\toprule
\textbf{Tool} & \textbf{ATE} & \textbf{95\% CI} & \textbf{Refuter} &
\shortstack{\textbf{New} \textbf{Effect}} &
\shortstack{\textbf{Est.}\textbf{Effect}} &
\textbf{$p$-value} \\
\midrule

\multirow{5}{*}{\sgrep} & \multirow{5}{*}{$0.109$} & \multirow{5}{*}{\shortstack{$[0.025,$\\$0.183]$}}
    & Random common cause     & $0.109$  & $0.109$ & $1.000$ \\
& & & Placebo treatment       & $-0.015$ & $0.109$ & $0.630$ \\
& & & Data subset             & $0.075$  & $0.109$ & $0.100$ \\
& & & Dummy outcome           & $-0.001$ & $0.000$ & $0.960$ \\
& & & Unobserved common cause & $0.250$  & $0.109$ & -- \\
\midrule

\multirow{5}{*}{\cql} & \multirow{5}{*}{$-0.057$} & \multirow{5}{*}{\shortstack{$[-0.110,$\\$-0.030]$}}
    & Random common cause     & $-0.057$ & $-0.057$ & $1.000$ \\
& & & Placebo treatment       & $0.008$  & $-0.057$ & $0.780$ \\
& & & Data subset             & $-0.074$ & $-0.057$ & $0.220$ \\
& & & Dummy outcome           & $-0.017$ & $0.000$  & $0.920$ \\
& & & Unobserved common cause & $0.628$  & $-0.057$ & -- \\
\midrule

\multirow{5}{*}{\ccrypt} & \multirow{5}{*}{$-0.310$} & \multirow{5}{*}{\shortstack{$[-0.394,$\\$-0.180]$}}
    & Random common cause     & $-0.310$ & $-0.310$ & $1.000$ \\
& & & Placebo treatment       & $-0.001$ & $-0.310$ & $0.930$ \\
& & & Data subset             & $-0.257$ & $-0.310$ & $0.140$ \\
& & & Dummy outcome           & $-0.013$ & $0.000$  & $0.910$ \\
& & & Unobserved common cause & $-0.122$ & $-0.310$ & -- \\
\midrule

\multirow{5}{*}{\cguard} & \multirow{5}{*}{$0.043$} & \multirow{5}{*}{\shortstack{$[-0.052,$\\$0.141]$}}
    & Random common cause     & $0.043$  & $0.043$ & $1.000$ \\
& & & Placebo treatment       & $0.083$  & $0.043$ & $0.020$ \\
& & & Data subset             & $-0.003$ & $0.043$ & $0.130$ \\
& & & Dummy outcome           & $0.001$  & $0.000$ & $0.970$ \\
& & & Unobserved common cause & $-0.161$ & $0.043$ & -- \\
\bottomrule
\end{tabular*}
}
\end{table}
\myparagraph{1. Overall Effect of Third-Party-Alert Reporting}
Table~\ref{tab:overall_refuter_results} reports the overall PSM estimates and APK-cluster bootstrap confidence intervals for the four detectors. 
The estimated effects differ substantially across tools.
\ccrypt{} exhibits the largest negative effect (ATE $=-0.310$), indicating that reporting third-party alerts decreases the probability that a reported alert is a true positive by 31.0 percentage points after adjustment.

\finding{{\bf Naive observational results may underestimate the causal effect} of the assumption, \ie while the baseline from a one-off experiment is  $0.216$ (about $22$\%) decrease in precision, the causal effect on reporting library code as well, in \ccrypt, is significantly higher, at $0.31$ precentage points.
%
}
%

\cql{} also shows a negative effect, although substantially smaller (ATE $=-0.057$). 
In contrast, \cguard{} and \sgrep{} exhibit small positive effects (ATEs of $0.043$ and $0.109$, respectively).


\finding{The assumption that reporting third-party library alerts decreases precision holds for \ccrypt and \cql, but not for \cguard and \sgrep. 
Here, {\bf tool design acts as an effect modifier}: the same reporting policy can produce different causal effects across tools because of differences in rule design, alert filtering, and implementation choices.}

\myparagraph{2. Effects Across Library Types}
%
To investigate the impact of including libary code at a more precise granularity, \ie of library types, we cluster libraries into four types using LibScout~\cite{bbd16}, \ie utilities, Android, Small libraries (combining the Social Media, Analytics, Advertising, and Cloud types as they had too few alerts each), and miscellaneous (which could not be classified by LibScout).
The treatment remains the same, \ie that of {\em including/reporting} alerts from libraries of a specific type, along with developer-written code, to estimate the causal effect.

Table~\ref{tab:library_estimator_comparison} in the Appendix reports the PSM, logistic-regression-adjustment, and doubly robust estimates for each tool library-type contrast.
We first describe the PSM estimates and subsequently examine their consistency with the other two estimators.

We find that the effect of third-party libraries on precision is not uniform: both the direction and magnitude of the effect vary across library families and \sast{}s.
For example, Utilities have the largest negative effect on precision for \ccrypt, with an ATE of $-0.375$; whereas 
 Android and Small Libraries lead to negligible ATE. 
This contrasts with \cguard, for which Android libraries have a small negative effect on precision, with an ATE of $-0.036$, but 
Utilities, Small Libraries, and miscellaneous libraries show positive effects, with ATEs of $0.181$, $0.044$, and $0.060$, respectively. 
This suggests that, unlike \ccrypt, including several categories of library alerts in \cguard does not reduce precision.
Similar variance is observed for other \sasts.

\finding{The {\bf causal effect} of reporting third-party library alerts is {\bf heterogeneous across both library families and \sasts.}}



\subsection{Results: Robustness and Consistency of the Estimated Effects}
\label{sec:eval-rpbustness}
%
We assess the credibility of the estimated effects using four complementary forms of evidence in order to address {\bf RQ3}. 
First, we examine the propensity-score overlap and covariate balance to determine whether the observed data provide comparable treated and control observations. 
Second, we examine agreement across three estimators: PSM (our primary estimator), logistic-regression adjustment, and doubly robust estimation.  
Third, we compare row-level and APK-cluster bootstrap intervals to account for dependence among alerts from the same application.
Finally, we evaluate the robustness of the estimated effects using five refuters for both the overall and the library-type causal analysis. 
Four refuters (random common cause, placebo treatment, data subset, and dummy outcome) assess stability under data perturbations and produce simulation-based p-values, while the unobserved common cause refuter evaluates sensitivity to a specified unmeasured-confounding scenario and does not produce a p-value.

\myparagraph{1. Overlap and Covariate Balance}
Before interpreting the effect estimates, we examine whether the treated and control groups contain sufficiently comparable observations.
Table~\ref{tab:overlap_balance_uncertainty} in the Appendix shows that only $0.08\%$--$2.33\%$ of observations fall outside common propensity-score support across the four tools, indicating that little extrapolation is required by the estimators. 
We next evaluate whether adjustment reduces differences (\ie the covariate balance) in the measured confounders (app popularity and APK size) using the maximum absolute standardized mean difference (SMD), where values below 0.1 conventionally indicate acceptable balance.
Table~\ref{tab:overlap_balance_uncertainty}  shows that the maximum absolute SMD ranges from $0.135$ to $0.375$ before adjustment, and after adjustment, decreases to $0.000$--$0.082$ with PSM, and $0.004$--$0.097$ with AIPW, indicating acceptable balance for both estimators across all four tools. 

\myparagraph{2. Comparison across 3 estimators} Table~\ref{tab:overall_estimator_comparison} compares the causal estimates produced by the three estimators, including PSM (our primary estimator).
For three of the four tools, all estimators agree on both the direction and approximate magnitude of the estimated causal effect.
The only exception is \cguard{}, where PSM estimates a positive effect while logistic-regression adjustment and doubly robust estimation both estimate negative effects.
Thus, unlike the remaining three tools, the overall causal effect for \cguard{} is inconclusive.

\myparagraph{3. APK-Dependent Uncertainty}
Because multiple alerts can originate from the same APK, treating alerts as independent may underestimate uncertainty.
We therefore compare a row bootstrap, which resamples individual alerts, with an APK-cluster bootstrap, which resamples complete APKs and retains their alerts together, thus preserving within-APK dependence.
After accounting for within-APK dependence, the PSM confidence interval remains positive for \sgrep and negative for \cql and \ccrypt.
In contrast, the \cguard interval includes zero, providing no conclusive evidence of either a precision increase or decrease.
Thus, accounting for APK-level depdendence preserves the directional conclusions for \sgrep, \cql, and \ccrypt, while reinforcing that the \cguard result is inconclusive.
%

\myparagraph{4. Refutation Tests: Robustness to Data Perturbations}
We first examine robustness under the four perturbation-based refuters (random common cause, placebo treatment, data subset, and dummy outcome).
The PSM results for the overall causal estimate are reported in Table~\ref{tab:overall_refuter_results}.

We interpret a refutation-test p-value below $0.05$ as evidence of statistically significant deviation from the original estimate, indicating sensitivity to that perturbation.
Across the four standard refuters, the overall causal estimates for \ccrypt, \sgrep, and \cql remain stable.
That is, the random common cause refuter leaves the estimated effects essentially unchanged, the placebo and dummy outcome refuters produce near-zero effects, and the data subset refuter produces effects close to the original estimates, with the values for data subsets indicating no statistically significant deviation from the original estimates. 
In contrast, \cguard is the only tool for which the overall PSM estimate is both inconsistent with the hypothesized direction of the effect and sensitive to the placebo refuter.

The results are similar for the library-type causal estimation, \ie largely stable under the four standard refuters, as seen in Table~\ref{tab:library_refuter_results} in the Appendix. 
The only significant data subset result occurs for \cguard Utilities: the estimate changes from $0.181$ to $0.125$, with a p-value of $0.030$. 
Although its magnitude changes, the estimate remains positive. 
The remaining library-type estimates do not show significant changes under the four standard refuters.

\finding{Standard perturbation-based refuters support the stability of most estimated causal effects, \ie for \ccrypt{}, \cql{}, and \sgrep{}. The only exception is \cguard{}, whose overall causal estimate is sensitive to the placebo refuter, and the and data-subset refuter for one library type, despite preserving the estimated direction.}



\myparagraph{5. Refutation Tests: Sensitivity to an Unobserved Common Cause}
We next evaluate robustness to omitted confounding using the unobserved common cause refuter, which simulates a binary confounder with treatment and outcome flip strengths of $0.05$.
Unlike the other refuters, this analysis does not produce p-values.
Under this perturbation, for the overall causal effect, the estimated effects for \sgrep{} ($0.109$ to $0.250$) and \ccrypt{} ($-0.310$ to $-0.122$) retain their original directions.
In contrast, \cql reverses from $-0.057$ to $0.628$, while \cguard reverses from $0.043$ to $-0.161$.
Thus, unlike for \sgrep{} and \ccrypt{}, the PSM estimates for \cql{} and \cguard{} are sensitive to the specified unobserved-confounding scenario.

For the libary-type causal estimation, the unobserved common cause refuter reveals substantially greater sensitivity.
Among the 16 library-type PSM estimates, 10 are in reverse direction under the specified scenario. 
The remaining six PSM contrasts retain their directions, although their magnitudes may change. 
Because this refuter does not produce p-values, these reversals should be interpreted as sensitivity to the specified unobserved-confounding strength, not as statistically significant refutation failures. 

\finding{While the estimated causal effects are generally stable under perturbation-based refutation analyses, sensitivity to the simulated unobserved-confounding scenario varies substantially across tools and library types. 
Thus, while supported under the measured causal model, the results should not be interpreted as robust to all unmeasured confounding.}


\section{Discussion and Takeaways}
\label{sec:discussion}

\sast{}s are a critical component of modern software development, and their design assumptions directly influence what vulnerabilities are ultimately detected in practice. 
Our systematic study reveals that \sast{}s rely on a diverse range of assumptions that are typically motivated by intuition, engineering experience, or limited empirical observations rather than explicit causal evidence ({\bf RQ1}), motivating \tools principled approach for evaluating their causal effects. 
As our evaluation demonstrates, these assumptions are neither uniformly valid nor uniformly invalid; instead, their effects depend on the underlying analysis, implementation, and software context ({\bf RQ2}). 
Our study also provides insight into both the strengths and limitations of applying causal inference to evaluate \sast{} assumptions, including its ability to distinguish robust causal conclusions from cases where the available evidence remains inconclusive ({\bf RQ3}).
We now synthesize our findings into broad themes that discuss the implications of these observations for the design, evaluation, and future development of \sast{}s.



\subsection{Security Assumptions are Pervasive, Diverse, and not Equally Testable}
Our systematic study identified 57 security assumptions spanning \detector rules, analysis heuristics, scope-reduction techniques, runtime optimizations, and end-user outcomes.
This diversity illustrates how heavily modern \sast{}s rely on assumptions to balance precision, recall, scalability, and usability, and provides a fertile ground for applying causal inference to evaluate their effects under explicit causal models. 
However, there is a caveat: {\em not all assumptions are equally amenable to causal evaluation in practice.}

While rule- and reporting-policy assumptions can often be operationalized through data-level treatments,  assumptions involving analysis techniques (\eg slicing or exploration depth) require method-level interventions that modify the analyzer itself.
Finally, assumptions involving outcomes such as developer productivity or satisfaction remain difficult to evaluate causally because suitable observational data are rarely available.

\begin{takeaway}{Takeaway 1}
The availability of diverse \sast design assumptions signifies both a problem and a research opportunity, particularly because many of these assumptions can be operationalized and evaluated systematically. 
However, the nature of their conditions and outcomes ultimately determines whether, and how easily, they can be operationalized for causal evaluation.

\end{takeaway}

\subsection{Causal Inference provides a Principled Alternative to Ad-hoc Validation}
Our evaluation demonstrates that causal inference provides substantially richer evidence than one-off observational comparisons by making treatments, confounders, and modeling assumptions explicit. 
For example, for \ccrypt, the adjusted causal effect was substantially larger than suggested by the raw precision comparison, illustrating the importance of controlling for confounding.
Thus, \tool is a principled improvement to the design and evaluation of \sasts. 

The practical deployment of causal inference for evaluating \sast assumptions, however, depends on engineering considerations. 
Data-level interventions are particularly attractive because existing alerts can often be reorganized into treatment conditions without modifying the analyzer, as in our reporting-policy evaluation. 
Their feasibility, however, depends on the availability of reliable labeled alert data; constructing our ground-truth dataset of 57,038 alerts required approximately 4.5 person-months of manual validation. 
In contrast, method-level interventions require modifying and repeatedly executing the analyzer for every configuration, and are substantially more expensive.

\begin{takeaway}{Takeaway 2}
When suitable treatments and ground-truth data can be constructed, causal inference provides a principled methodology for evaluating \sast{} assumptions that goes beyond unadjusted empirical observations. 
Data-level interventions make many practical evaluations feasible, whereas method-level interventions and ground-truth construction remain the primary engineering bottlenecks.
\end{takeaway}

\subsection{The Validity of Security Assumptions is Fundamentally Context Dependent}
A central observation from our evaluation is that security assumptions cannot be treated as universally valid.
That is, we observe how the same reporting-policy assumption produces substantially different causal effects across \ccrypt{}, \cql{}, \cguard{}, and \sgrep{}, demonstrating that tool design acts as an effect modifier.
Furthermore, the estimated effects varied across library families, indicating that implementation choices and software context jointly influence the causal behavior of a given assumption.
That is, there is no guarantee that a broadly accepted assumption will transfer across SASTs, fundamentally changing how future \sast{}s should interpret and build upon prior empirical findings.


\begin{takeaway}{Takeaway 3}
Security assumptions are context dependent. 
Their causal effects depend on the specific \sast{}, its implementation, and the software ecosystem under evaluation. 
Thus, assumptions should be evaluated in the context of the target \sast rather than inherited from prior tools.
\end{takeaway}

\subsection{Causal inference is Valuable even When it Produces Inconclusive Answers}
One of the most important methodological observations from our study is that causal inference does not always produce a definitive answer.
For \cguard{}, three independent analyses point to the same conclusion, \ie (1) the three estimators disagree on the direction of the causal effect, (2) the APK-cluster bootstrap confidence interval indicates that the estimated effect is not distinguishable from no effect after accounting for within-APK dependence, and finally, (3) the estimated effect is sensitive to both placebo perturbations and simulated unobserved confounding.
Rather than representing a limitation of causal inference, these observations demonstrate one of its principal strengths: the methodology explicitly distinguishes between assumptions that are well supported by the available evidence and those for which reliable causal conclusions cannot (yet) be drawn.

\begin{takeaway}{Takeaway 4}
An inconclusive causal result is itself a scientifically meaningful outcome, distinguishing assumptions that are well supported by the available evidence from those that are not.
\end{takeaway}



\section{Data Availability}
\label{sec:data_availability}

Our replication package is publicly available in an anonymized repository~\cite{anonymousgithub}.

\bibliographystyle{ACM-Reference-Format}
\bibliography{reference,phone,iot,mutation-references,mutation_security,cryptoapi,compliance,taint}

\appendix



\section{Analysis of Assumptions - SLR Search Query}
\label{app:assumptions}
We used the following search query for finding our initial list of papers in identified sources for our systematic literature review:

\noindent {\em ("static analysis" OR "SAST" OR "static application security testing")
AND
("taint analysis" OR "taint tracking" OR "data leak detection" OR "crypto API misuse" OR "cryptographic API misuse" OR "API misuse" OR "vulnerability detection")}


\label{app:eval}
\begin{table*}[!h]
\centering
\scriptsize
\caption{Propensity-score overlap, covariate balance, and bootstrap uncertainty diagnostics. The balance columns report the maximum absolute standardized mean difference across the adjustment variables. Values below $0.1$ indicate acceptable covariate balance.}
\setlength{\tabcolsep}{4pt}
\renewcommand{\arraystretch}{1.25}
\begin{tabular}{|l|r|r|r|r|c|c|}
\hline
\textbf{Tool} &
\shortstack{\textbf{Outside Common}\\\textbf{Support (\%)}} &
\shortstack{\textbf{Max. $|\mathrm{SMD}|$}\\\textbf{Before Adjustment}} &
\shortstack{\textbf{Max. $|\mathrm{SMD}|$}\\\textbf{After PSM}} &
\shortstack{\textbf{Max. $|\mathrm{SMD}|$}\\\textbf{After AIPW}} &
\shortstack{\textbf{Row-Bootstrap}\\\textbf{PSM 95\% CI}} &
\shortstack{\textbf{APK-Cluster}\\\textbf{PSM 95\% CI}} \\
\hline

\sgrep
& $0.08$
& $0.375$
& $0.000$
& $0.030$
& $[0.023,\,0.129]$
& $[0.025,\,0.183]$ \\
\hline

\cql
& $1.03$
& $0.135$
& $0.082$
& $0.004$
& $[-0.095,\,-0.023]$
& $[-0.110,\,-0.030]$ \\
\hline

\ccrypt
& $2.33$
& $0.178$
& $0.078$
& $0.010$
& $[-0.348,\,-0.171]$
& $[-0.394,\,-0.180]$ \\
\hline

\cguard
& $0.18$
& $0.169$
& $0.013$
& $0.097$
& $[-0.067,\,0.076]$
& $[-0.052,\,0.141]$ \\
\hline

\end{tabular}
\label{tab:overlap_balance_uncertainty}
\end{table*}

\begin{table}[t]
\centering
\scriptsize
\setlength{\tabcolsep}{2.4pt}
\renewcommand{\arraystretch}{1.30}
\begin{tabular}{|l|p{0.52\columnwidth}|r|r|}
\hline
\textbf{Tool} & \textbf{Setting} & \textbf{\# of Alerts} & \textbf{Prec.} \\
\hline

\multirow{9}{*}{\sgrep}
& \textbf{Overall (all alerts)} & $\textbf{14590}$ & $\textbf{0.6743}$ \\ \cline{2-4}
& Developer-written only (is\_third\_party=0) & $1760$ & $0.5494$ \\ \cline{2-4}
& Third-party only (is\_third\_party=1) & $12830$ & $0.6914$ \\ \cline{2-4}
& Balanced A (downsampled) - Combined & $3520$ & $0.6227$ \\ \cline{2-4}
& Balanced A - Third-party (downsampled) & $1760$ & $0.6960$ \\ \cline{2-4}
& Balanced B (upsampled) - Combined & $25660$ & $0.6208$ \\ \cline{2-4}
& Balanced B - Developer-written (upsampled) & $12830$ & $0.5501$ \\ 
\hline

\multirow{9}{*}{\cql}
& \textbf{Overall (all alerts)} & $\textbf{21501}$ & $\textbf{0.8887}$ \\ \cline{2-4}
& Developer-written only (is\_third\_party=0) & $554$ & $0.9513$ \\ \cline{2-4}
& Third-party only (is\_third\_party=1) & $20947$ & $0.8870$ \\ \cline{2-4}
& Balanced A (downsampled) - Combined & $1108$ & $0.9233$ \\ \cline{2-4}
& Balanced A - Third-party (downsampled) & $554$ & $0.8953$ \\ \cline{2-4}
& Balanced B (upsampled) - Combined & $41894$ & $0.9193$ \\ \cline{2-4}
& Balanced B - Developer-written (upsampled) & $20947$ & $0.9515$ \\ 
\hline

\multirow{9}{*}{\ccrypt}
& \textbf{Overall (all alerts)} & $\textbf{5675}$ & $\textbf{0.4846}$ \\ \cline{2-4}
& Developer-written only (is\_third\_party=0) & $498$ & $0.7008$ \\ \cline{2-4}
& Third-party only (is\_third\_party=1) & $5177$ & $0.4638$ \\ \cline{2-4}
& Balanced A (downsampled) - Combined & $996$ & $0.5914$ \\ \cline{2-4}
& Balanced A - Third-party (downsampled) & $498$ & $0.4819$ \\ \cline{2-4}
& Balanced B (upsampled) - Combined & $10354$ & $0.5863$ \\ \cline{2-4}
& Balanced B - Developer-written (upsampled) & $5177$ & $0.7089$ \\ 
\hline

\multirow{9}{*}{\cguard}
& \textbf{Overall (all alerts)} & $\textbf{15272}$ & $\textbf{0.4817}$ \\ \cline{2-4}
& Developer-written only (is\_third\_party=0) & $1318$ & $0.5129$ \\ \cline{2-4}
& Third-party only (is\_third\_party=1) & $13954$ & $0.4788$ \\ \cline{2-4}
& Balanced A (downsampled) - Combined & $2636$ & $0.4981$ \\ \cline{2-4}
& Balanced A - Third-party (downsampled) & $1318$ & $0.4833$ \\ \cline{2-4}
& Balanced B (upsampled) - Combined & $27908$ & $0.4962$ \\ \cline{2-4}
& Balanced B - Developer-written (upsampled) & $13954$ & $0.5135$ \\ 
\hline
\end{tabular}
\caption{Precision Baselines Before Causal Adjustment. Balanced A = Downsample alerts from third-party lib to match alerts from developer-written count; Balanced B = Upsample alerts from developer-written to match alerts from third-party count}
\label{tab:raw_precision_results}
\end{table}

\begin{table}[!t]
\centering
\scriptsize
\caption{Comparison of PSM, logistic regression adjustment, and doubly robust estimates.}
\setlength{\tabcolsep}{10pt}
\renewcommand{\arraystretch}{1.25}
\begin{tabular}{|l|r|r|r|c|}
\hline
\textbf{Tool} & \textbf{PSM ATE} & \textbf{Logistic ATE} &
\textbf{Doubly Robust ATE} & \textbf{Same Direction?} \\
\hline
\sgrep & $0.109$ & $0.122$ & $0.116$ & Yes \\
\hline
\cql & $-0.057$ & $-0.064$ & $-0.060$ & Yes \\
\hline
\ccrypt & $-0.310$ & $-0.223$ & $-0.238$ & Yes \\
\hline
\cguard & $0.043$ & $-0.030$ & $-0.023$ & No / mixed \\
\hline
\end{tabular}
\label{tab:overall_estimator_comparison}
\end{table}

\begin{table}[t]
\centering
\scriptsize
\caption{Comparison of PSM, logistic regression adjustment, and doubly robust estimates.}
\setlength{\tabcolsep}{4pt}
\renewcommand{\arraystretch}{1.15}
\begin{tabular}{|l|l|r|r|r|c|}
\hline
\textbf{Tool} & \textbf{Library Type} & \textbf{PSM ATE} &
\textbf{Logistic ATE} & \textbf{Doubly Robust ATE} &
\textbf{Same Direction?} \\
\hline

\multirow{4}{*}{\sgrep}
& Android & $0.094$ & $0.103$ & $0.101$ & Yes \\ \cline{2-6}
& Small Libraries & $0.055$ & $0.036$ & $0.036$ & Yes \\ \cline{2-6}
& Utilities & $0.029$ & $0.045$ & $0.045$ & Yes \\ \cline{2-6}
& Miscellaneous & $0.079$ & $0.103$ & $0.103$ & Yes \\
\hline

\multirow{4}{*}{\cql}
& Android & $-0.014$ & $-0.021$ & $-0.015$ & Yes \\ \cline{2-6}
& Small Libraries & $-0.006$ & $0.005$ & $0.005$ & No / mixed \\ \cline{2-6}
& Utilities & $-0.043$ & $-0.047$ & $-0.049$ & Yes \\ \cline{2-6}
& Miscellaneous & $-0.132$ & $-0.134$ & $-0.130$ & Yes \\
\hline

\multirow{4}{*}{\ccrypt}
& Android & $-0.048$ & $0.030$ & $0.030$ & No / mixed \\ \cline{2-6}
& Small Libraries & $-0.031$ & $0.067$ & $0.067$ & No / mixed \\ \cline{2-6}
& Utilities & $-0.375$ & $-0.338$ & $-0.345$ & Yes \\ \cline{2-6}
& Miscellaneous & $-0.112$ & $-0.087$ & $-0.105$ & Yes \\
\hline

\multirow{4}{*}{\cguard}
& Android & $-0.036$ & $-0.131$ & $-0.136$ & Yes \\ \cline{2-6}
& Small Libraries & $0.044$ & $0.072$ & $0.075$ & Yes \\ \cline{2-6}
& Utilities & $0.181$ & $0.143$ & $0.148$ & Yes \\ \cline{2-6}
& Miscellaneous & $0.060$ & $-0.023$ & $-0.012$ & No / mixed \\
\hline

\end{tabular}
\label{tab:library_estimator_comparison}
\end{table}






\begin{table*}[t]
\centering
\scriptsize
\caption{PSM causal estimates and refutation results. Each refuter cell reports the new effect with its p-value in parentheses. The unobserved-common-cause sensitivity analysis does not produce a p-value. Here, RCC=Random Common Cause, PTR=Placebo Treatment Refuter, DSR=Data Subset Refuter, DOR=Dummy Outcome Refuter and UCC=Unobserved Common Cause}
\setlength{\tabcolsep}{2pt}
\renewcommand{\arraystretch}{1.25}
\begin{tabular}{|l|l|r|c|c|c|c|c|}
\hline
\textbf{Tool} & \textbf{Library} & \textbf{ATE} &
\textbf{RCC} &
\textbf{PTR} &
\textbf{DSR} &
\textbf{DOR} &
\textbf{UCC} \\
\hline

\multirow{4}{*}{\sgrep}
& Android & $0.094$ & $0.094\,(1.000)$ & $0.003\,(0.950)$ &
$0.070\,(0.140)$ & $-0.002\,(0.940)$ & $0.168\,$ \\ \cline{2-8}
& Misc. & $0.079$ & $0.079\,(1.000)$ & $0.007\,(0.780)$ &
$0.060\,(0.410)$ & $0.009\,(0.880)$ & $0.176\,$ \\ \cline{2-8}
& Utilities & $0.029$ & $0.029\,(1.000)$ & $0.001\,(0.990)$ &
$0.004\,(0.250)$ & $0.001\,(0.960)$ & $-0.011\,$ \\ \cline{2-8}
& Small Lib. & $0.055$ & $0.055\,(1.000)$ & $0.000\,(0.960)$ &
$0.025\,(0.120)$ & $-0.011\,(0.900)$ & $-0.001\,$ \\
\hline

\multirow{4}{*}{\cql}
& Android & $-0.014$ & $-0.014\,(1.000)$ & $0.004\,(0.840)$ &
$-0.018\,(0.760)$ & $-0.002\,(0.920)$ & $0.644\,$ \\ \cline{2-8}
& Misc. & $-0.132$ & $-0.132\,(1.000)$ & $0.005\,(0.910)$ &
$-0.134\,(0.920)$ & $0.002\,(0.990)$ & $0.399\,$ \\ \cline{2-8}
& Utilities & $-0.043$ & $-0.043\,(1.000)$ & $-0.003\,(0.920)$ &
$-0.052\,(0.420)$ & $0.008\,(0.990)$ & $0.434\,$ \\ \cline{2-8}
& Small Lib. & $-0.006$ & $-0.006\,(1.000)$ & $-0.002\,(0.815)$ &
$0.006\,(0.490)$ & $0.004\,(0.990)$ & $0.166\,$ \\
\hline

\multirow{4}{*}{\ccrypt}
& Misc. & $-0.112$ & $-0.112\,(1.000)$ & $-0.001\,(0.915)$ &
$-0.093\,(0.650)$ & $-0.015\,(0.860)$ & $0.069\,$ \\ \cline{2-8}
& Utilities & $-0.375$ & $-0.375\,(1.000)$ & $-0.011\,(0.735)$ &
$-0.362\,(0.760)$ & $0.003\,(0.940)$ & $-0.266\,$ \\ \cline{2-8}
& Android & $-0.048$ & $-0.048\,(1.000)$ & $0.005\,(0.920)$ &
$0.013\,(0.130)$ & $-0.004\,(0.940)$ & $0.078\,$ \\ \cline{2-8}
& Small Lib. & $-0.031$ & $-0.031\,(1.000)$ & $0.001\,(0.930)$ &
$0.028\,(0.140)$ & $-0.002\,(0.970)$ & $0.083\,$ \\
\hline

\multirow{4}{*}{\cguard}
& Misc. & $0.060$ & $0.060\,(1.000)$ & $0.024\,(0.545)$ &
$0.022\,(0.170)$ & $-0.010\,(0.840)$ & $-0.053\,$ \\ \cline{2-8}
& Android & $-0.036$ & $-0.036\,(1.000)$ & $0.033\,(0.170)$ &
$-0.074\,(0.120)$ & $0.003\,(0.960)$ & $-0.175\,$ \\ \cline{2-8}
& Utilities & $0.181$ & $0.181\,(1.000)$ & $-0.010\,(0.630)$ &
$0.125\,(0.030)$ & $-0.005\,(0.980)$ & $0.139\,$ \\ \cline{2-8}
& Small Lib. & $0.044$ & $0.044\,(1.000)$ & $0.007\,(0.885)$ &
$0.065\,(0.300)$ & $0.004\,(0.950)$ & $0.051\,$ \\
\hline

\end{tabular}
\label{tab:library_refuter_results}
\end{table*}

\end{document}